\documentclass[12pt]{iopart}
\def\ns{\mathcal{NS}}
\def\l{\mathcal{L}}
\def\Q{\mathcal{Q}}
\def\g{\mathcal{G}}
\def\e{\mathcal{E}}
\def\u{\mathcal{U}}
\def\r{\mathcal{R}}
\def\k{\mathcal{K}}
\def\o{\mathcal{O}}
\def\D{\mathcal{D}}
\def\p{\textbf{p}}
\def\q{\textbf{q}}
\def\d{\textbf{d}}
\def\s{\textbf{s}}
\def\L{\textbf{l}}

\def\co{\mathrm{co}}

\usepackage{iopams}

\begin{document}

\title[On nonlocality as a resource theory and nonlocality measures]{On nonlocality as a resource theory and nonlocality measures}

\author{Julio I de Vicente}

\address{Departamento de Matem\'aticas, Universidad Carlos III
de Madrid, Avda.\ de la Universidad 30, 28911 Legan\'es, Madrid,
Spain} \ead{jdvicent@math.uc3m.es}
\begin{abstract}
With the advent of device independent quantum information processing, nonlocality is nowadays regarded as a resource to implement various tasks. On the analogy of entanglement theory we approach nonlocality from this perspective. In order to do so, we analyze in full detail the operations that can be implemented in this scenario and under which nonlocality cannot increase. This provides a theoretical ground to study how to order and quantify nonlocal behaviours. Finally, we review several nonlocality measures and discuss their validity from this point of view.
\end{abstract}

%Uncomment for PACS numbers title message
% Keywords required only for MST, PB, PMB, PM, JOA, JOB?
%\vspace{2pc}
%\noindent{\it Keywords}: Article preparation, IOP journals
% Uncomment for Submitted to journal title message
%\submitto{\JPA}
% Comment out if separate title page not required
%\maketitle

\section{Introduction}

Bell's theorem states that the predictions of quantum mechanics are not compatible with any local realistic theory (i.\ e.\ any local hidden variable model) \cite{bell}. This result has deep conceptual implications and constitutes a milestone in our understanding of quantum mechanics. However, 50 years after its discovery, the relevance of quantum nonlocality goes beyond foundational issues. The development of quantum information science \cite{nielsenchuang} in the last decades has given rise to many efforts towards a better understanding on the possibilities and limitations of quantum theory. Not surprisingly, the insight behind Bell's theorem has played an important role in helping to develop this task.

It has been a standard claim that entanglement is a key resource in the applications of quantum information theory. This has led to the development of entanglement theory, which not only aims at its characterization but also at understanding its manipulation and quantification \cite{entreviews}. In this theory entanglement is a resource when state manipulation is restricted to a certain class of operations known as local operations and classical communication (LOCC). Thus, the basic law of entanglement theory is that entanglement cannot increase under LOCC and any entanglement measure must satisfy this principle.

Interestingly, although the relevance of entanglement is unquestionable, it has been noticed that, depending on the scenario, other properties of quantum theory might play the crucial role for the quantum-over-classical advantage. Given the conceptual elegance of entanglement theory, this has led to the development of other resource theories such as frame alignment \cite{frameness} or stabilizer computation \cite{magic}. The emergent subfield of device independent quantum information processing (DIQIP) is another instance of this situation. In this setting all devices are regarded as black boxes. Hence, no assumption is made on the actual measurements being carried out nor on the state being measured. Only measurement statistics are accounted. This scenario is sometimes more convenient from the point of view of experimental implementation and protocols that can be carried out in this way include quantum key distribution \cite{qkd}, randomness generation \cite{re} and dimensionality (see e.\ g.\ \cite{dim,diment}) and entanglement (see e.\ g.\ \cite{ent,diment}) certification. Naturally, the key resource for DIQIP is quantum nonlocality. This motivates the construction of a resource theory of nonlocality. Since it is now well-known that nonlocality and entanglement are two different concepts \cite{werner}, although this theory can be built on the analogy of entanglement theory, it must necessarily be a different one.

This approach to nonlocality was considered in \cite{barrett} and several authors have pursued different directions in this context (see e.\ g.\ \cite{JM,BP,FW} for the study of interconversions among several copies of nonlocal resources, \cite{distillation} for the distillation of several copies of a nonlocal behaviour into a more nonlocal one and Sec.\ V of the review \cite{nlreview}). Of particular relevance in this context is the recent work \cite{wccpi}, which, analogously to LOCC for entanglement, establishes the operations under which nonlocality cannot increase and uses this to find a proper definition of genuinely multipartite nonlocality (see also \cite{gmnl}). However, independently of this several measures of nonlocality arising from different motivations have been considered in the literature over the last decades (see Sec.\ V below). The main aim of the present article is to present a comprehensive analysis of nonlocality as a resource theory for single-copy manipulations on the analogy of the same setting in entanglement theory. Following \cite{wccpi} and previous references, we will carefully describe in full mathematical detail the operations that can be implemented in the DIQIP scenario. Under these transformations nonlocality cannot increase, thus inducing a natural ordering. We will then reconsider several nonlocality measures from this perspective to asses whether they meet this condition and are therefore meaningful quantifiers. It is the author's hope that the present work will help to establish further nonlocality as a resource theory and to have a systematic and rigorous treatment of nonlocality measures.

\section{The DIQIP scenario}

For simplicity, in this paper we will only consider the bipartite case. There are two parties, Alice ($A$) and Bob ($B$), which share a state from which they have no information. Each of them has a device which performs measurements on their share of the state but its internal functioning is unknown to them. They only know that each of them can choose different input measurements labeled by $x\in\{1,2,\cdots,m_A\}$ and $y\in\{1,2,\cdots,m_B\}$ respectively. For every input measurement the boxes can output different results labeled by $a\in\{1,2,\cdots,d_A\}$ and $b\in\{1,2,\cdots,d_B\}$ respectively\footnote{There is in principle the possibility of allowing the cardinality of the set of outputs to depend on the input. However, at the end of the day, one can always consider that all inputs have the same number of possible outputs but some of them never occur.}. Given this setting, Alice and Bob can have access to the joint probability distribution of obtaining the outputs $(a,b)$ given the choice of inputs $(x,y)$, which we denote by $P(ab|xy)$. This list of $d_Ad_Bm_Am_B$ numbers, which is usually termed as behaviour, cannot take arbitrary values. Of course, the fact that they are probabilities imposes that $P(ab|xy)\geq0$ $\forall a,b,x,y$ and that $\sum_{a,b}P(ab|xy)=1$ $\forall x,y$. Moreover, in order to be consistent with special relativity, the no-signaling conditions need to be fulfilled
\begin{eqnarray}
\sum_bP(ab|xy)=\sum_bP(ab|xy'),\quad \forall a,x,y,y',\\
\sum_aP(ab|xy)=\sum_aP(ab|x'y),\quad \forall b,x,x',y.
\end{eqnarray}
This amounts to the fact that the local marginal probabilities for each party $P(a|x)$ and $P(b|y)$ do not depend on the other's party choice of measurement. This is because the parties might be space-like separated and this would otherwise allow to signal to one another. From the information-theoretic perspective, this should be understood as that once the measurement phase has begun the parties are not allowed to communicate with each other.

Thus, we will only consider the set of behaviours satisfying the aforementioned conditions, which we will refer to as no-signaling behaviours and will denote by $\mathcal{NS}$. This is the main object of study and the (possible) resource in DIQIP in the same way as (entangled) quantum states are in entanglement theory. The parties are provided with a state and measurement devices with the promise that they operate under a given behaviour in $\mathcal{NS}$. They can use this to implement some protocol but once they have used the devices to measure, the resource is destroyed. This is exactly analogous to the case of parties which have been provided with a given quantum state in quantum information theory. Here, we will only consider single-copy manipulations of the resource. This means that although the parties might be provided with several instances of the resource behaviour, they cannot access them jointly.

\section{Is a behaviour nonlocal?}

A no-signaling behaviour is said to be local if there exists a probability measure $\mu(\lambda)$ over some sample space $\Lambda$ such that
\begin{equation}\label{local}
P(ab|xy)=\int_\Lambda d\mu(\lambda)P(a|x,\lambda)P(b|y,\lambda),\quad \forall a,b,x,y,
\end{equation}
where $P(a|x,\lambda)\geq0$ $\forall a,x,\lambda$ and $P(b|y,\lambda)\geq0$ $\forall b,y,\lambda$ are probability distributions over the outputs $\{a\}$ and $\{b\}$ respectively, i.\ e.\ $\sum_aP(a|x,\lambda)=\sum_bP(b|y,\lambda)=1$ $\forall x,y$ and $\forall \lambda\in\Lambda$. The local subset of $\ns$ of all behaviours that can be written as in Eq.\ (\ref{local}) is denoted by $\l$. From the point of view of DIQIP, this is the subset of useless objects, i.\ e.\ those that do not constitute a resource. This is analogous to the subset of separable states in entanglement theory.

Although all behaviours in $\ns$ are compatible with the no-signaling principle, this does not mean that they can be observed in nature. Our best description (so far) is given by quantum theory. Thus, the set $\Q$ of all behaviours allowed by this theory is of particular relevance. %, i.\ e.\ those for which there exist a unit trace positive semidefinite operator $\rho$ over some Hilbert space $H_A\otimes H_B$ and positive semidefinite operators $\{A_a^x\}$ ($\{B_b^y\}$) acting on $H_A$ ($H_B$) fulfilling $\sum_aA_a^x=\one_A$ $\forall x$ and $\sum_bB_b^y=\one_B$ $\forall y$ such that
%\begin{equation}\label{quantum}
%P(ab|xy)=\tr(\rho A_a^x\otimes B_b^y).
%\end{equation}
 Remarkably, we have that
%\begin{equation}
$\l\subset\Q\subset\ns$,
%\end{equation}
 where all inclusions are strict. The first one is the content of Bell's theorem \cite{bell}, while the second one was observed by Popescu and Rohrlich \cite{PR} and Tsirelson \cite{Tsi}. This means that, although not exploring the full possibilities of no-signaling behaviours, quantum mechanics provides a way to obtain nonlocal behaviours and DIQIP is thus possible in practice.

This said, the first natural question for the theory of nonlocality is of course, given a behaviour in $\ns$, does it belong to $\l$ or not? That is, does it constitute a useful resource or not? This question is equivalent to the separability problem in entanglement theory (is a given quantum state entangled or not?). From now on, following standard notation we will arrange the numbers $P(ab|xy)$ of a behaviour into a point $\p\in\mathbb{R}^{d_Ad_Bm_Am_B}$. A local behaviour is said to be deterministic if an outcome is obtained with certainty for every choice of measurement for each party, i.\ e.\
\begin{equation}\label{deterministic}
P(ab|xy)=\delta_{a,f(x)}\delta_{b,g(y)},
\end{equation}
where $f$ ($g$) is any function mapping the elements of the input alphabet $\{1,2,\cdots,m_A\}$ ($\{1,2,\cdots,m_B\}$) to an element of the output alphabet $\{1,2,\cdots,d_A\}$ ($\{1,2,\cdots,d_B\}$). For a given setting, there are $d_A^{m_A}d_B^{m_B}$ different local deterministic behaviours which will be denoted by $\{\d_i\}$ in vector form. Fortunately, it turns out that the definition of the local subset $\l$ given by Eq.\ (\ref{local}) is equivalent to \cite{fine,nlreview}
\begin{equation}\label{localdet}
\p=\sum_ip_i\d_i,
\end{equation}
where $\{p_i\}$ is a (discrete) probability distribution. Hence, this immediately implies that $\l$ is a compact convex set with a finite number of extremal points (vertices), that is, a convex polytope. Therefore, to check whether a behaviour is local or not reduces to checking whether a point belongs to a given polytope or not. A polytope can be described by its vertices or by its facets% (i.e. by the intersection of a finite number of half-spaces)
, that is by a finite set of linear constraints of the form
\begin{equation}\label{bellineq}
\s\cdot\p\leq S\in\mathbb{R},
\end{equation}
which is nothing but a Bell-type inequality. Thus, the problem reduces to obtaining all facet Bell inequalities for every possible setting $\{m_A,m_B,d_A,d_B\}$. There are several useful tools to approach such a problem (see the review \cite{nlreview} and references therein). Here, we will only briefly mention that to check whether a behaviour can be written as in Eq.\ (\ref{localdet}) can be cast as a linear programming problem, its dual form yielding a Bell inequality that is violated when $\p\notin\l$. Therefore, to check whether a given behaviour is local or not can in principle be easily solved. However, it must be stressed that since the number of local deterministic points increases exponentially with the number of inputs, this problem becomes computationally intractable as this number grows.

\section{Is a behaviour more nonlocal than another one?}

Once the question whether an object is a resource or not is answered, the next step in a resource theory is to clarify which resources are more useful than others. This amounts to ordering the set of elements which constitute a resource. That is, we are looking for a binary relation $\succ$ to compare different elements in the set $\ns$ such that if $\p_1\succ\p_2$ holds then $\p_1$ is not less nonlocal than $\p_2$, and, hence, more (or equally) useful as a resource for DIQIP. Of course, there must be some operational meaning behind the ordering relation $\succ$. To do so, one considers that the parties are limited in the operations they can implement in such a way that they can only prepare non-resource objects (i.\ e.\ local behaviours in our case). This justifies the name ``resource'' to any object that grants the parties the power to overcome the limitations of what is achievable only within the set of allowed operations. For instance, in the case of entanglement theory the allowed set of operations is LOCC, from which parties can only prepare separable (i.\ e.\ non-entangled states). Entangled states can only be given to them. Now, if one entangled state $|\phi\rangle$ can be obtained from another entangled state $|\psi\rangle$ by LOCC, then $|\phi\rangle$ cannot be more entangled than $|\psi\rangle$. This is most reasonable as any task that can be achieved by the parties with $|\phi\rangle$ in this context can then also be achieved with $|\psi\rangle$, but not necessarily the other way around. Therefore, $|\psi\rangle$ is more (or at least equally) useful than $|\phi\rangle$ and, hence, more (or at least equally) entangled. Thus, the question relevant in our context is what operations are ``given for free'' to the parties in a DIQIP scenario. This has been recently addressed in \cite{wccpi} and the set of allowed operations has been termed as WCCPI (wiring and classical communication prior to the inputs). This comes from the fact that communication among the parties is only allowed outside the measurement phase, i.\ e.\ before the inputs are chosen and after the outcomes are recorded in order to be consistent with the no-signaling condition. The allowed classical communication as well as shared randomness are unlimited. Moreover, locally the parties have unrestricted power, i.\ e.\ they can prepare any marginal probability distribution. In the following I analyze in full detail and state mathematically which operations over the set $\ns$ constitute WCCPI.

\subsection{The allowed set of operations in the DIQIP scenario}

Below we present the different kinds of operations that can be implemented. Once all of them have been introduced, the more general transformation that can can be achieved is a composition of all of them. However, it will be convenient to keep in mind that compositions are non-trivial because mixing is allowed. That is, if certain operations allow to transform $\p_1$ into $\p_2$ and other allowed operations $\p_1$ into $\p_3$, then the parties, when given $\p_1$, can use the unlimited shared randomness and the communication before the inputs to agree to prepare $\p_2$ or $\p_3$ with certain probability and, thus, $\p_1$ can also be transformed to $\lambda\p_2+(1-\lambda)\p_3$ $\forall\lambda\in[0,1]$.

\subsubsection{Relabelings}

Before establishing which operations cannot increase nonlocality, one should clarify which operations do not change nonlocality, i.\ e.\ those under which the behaviours are equivalent. This has to correspond to the set of operations which can be reversed within the set of allowed operations. This would be analogous in entanglement theory to local unitary operations which are invertible LOCC transformations and amount to mere local basis rotations. In the DIQIP scenario, it is clear that the labeling of the inputs as well as the labeling of the outputs for each input are irrelevant and completely up to each party. Thus, one should regard behaviours related in this way as equal and the ordering $\succ$ with respect to these equivalence classes. That is, if $P_1(ab|xy)=P_2(\pi_x^1(a)\pi^2_y(b)|\pi^3(x)\pi^4(y))$, where $\pi^3$ ($\pi^4$) is any function mapping Alice's (Bob's) input alphabet $\{1,2,\cdots,m_A\}$ ($\{1,2,\cdots,m_B\}$) to a permutation of it and $\{\pi_x^1\}$ ($\{\pi^2_y\}$) any functions mapping Alice's (Bob's) output alphabet $\{1,2,\cdots,d_A\}$ ($\{1,2,\cdots,d_B\}$) to a permutation of it, then both behaviours are equally nonlocal. We will denote any relabeling of the form above (including the identity operation) by $\r(\p)$. Notice, indeed, that the inverse of any relabeling operation is another relabeling operation and, hence, if $\p_1=\r(\p_2)$ then $\p_2=\r'(\p_1)$.
\begin{itemize}
\item[(P1)] Given $\p_1,\p_2\in\ns$, if $\p_1=\r(\p_2)$ for some relabeling operation $\r$, then $\p_1$ and $\p_2$ are equally nonlocal ($\p_1\succ\p_2$ and $\p_2\succ\p_1$).
\end{itemize}
As explained above, the parties can then mix their given resource with any relabeling of it and, thus, it should hold that $\p\succ\sum_ip_i\r_i(\p)$ for any convex weights $\{p_i\}$ (i.\ e.\ $\sum_ip_i=1$ and $p_i\geq0$ $\forall i$) and relabeling operations $\{\r_i\}$. That is, $\p\succ\co(\{\r_i(\p)\})$, where $\co(\{x_i\})$ denotes any element in the convex hull of $\{x_i\}$.

\subsubsection{Mixing with local behaviours}

Since the parties have unlimited access to shared randomness and full power to prepare any marginal probability distribution, they can locally prepare any independent behaviours $p(a|x,\lambda)p(b|y,\lambda)$ and then use the classical communication allowed before the choice of inputs to correlate them to obtain any behaviour of the form (\ref{local}), i.\ e.\ any local behaviour. Moreover, as already discussed, when given a nonlocal resource behaviour, they can mix it with any local behaviour they have prepared before carrying out the measurements by using again shared randomness and the allowed classical communication at this stage.
\begin{itemize}
\item[(P2)] Given $\p_1,\p_2\in\ns$, if $\p_2=\lambda\p_1+(1-\lambda)\L$ for any $\lambda\in[0,1]$ and any $\L\in\l$, then $\p_1\succ\p_2$.
\end{itemize}

\subsubsection{Output operations}

Thus, the allowed set of operations basically gives the parties the power to mix their resource behaviour with any local behaviour. However, besides relabeling, they can also act on the input and output alphabets locally. The first that one can consider is that one party, say Alice, can choose to regard a certain subset of the outputs (possibly conditioned on the input) to be exactly the same event. Let $\{\mathcal{S}^A_x\}$ be (possibly different depending on the input $x$) subsets of Alice's output alphabet. Then Alice can map the initial behaviour $P$ into a new one in which all the outputs corresponding to $\{\mathcal{S}^A_x\}$ are considered as one single output and occur now with probability $\sum_{a\in\mathcal{S}^A_x}P(ab|xy)$. This would in principle change the setting as the new behaviour would have less possible outputs. Although this gives the possibility to compare behaviours over different settings, this can also be used to compare behaviours over the same setting which is what we will assume unless otherwise stated. This is because we can regard the new behaviour to have the same number of outputs as the original one but some of them occur now with probability zero. That is, $P$ is mapped to $P'$ in the same setting according to
\begin{equation}\label{ocg}
P'(ab|xy)=\left\{\begin{array}{cc}
                  P(ab|xy) & a\notin\mathcal{S}^A_x\\
                  \sum_{a'\in\mathcal{S}^A_x}P(a'b|xy) & a=a_x\in\mathcal{S}^A_x \\
                  0 & a\neq a_x,a\in\mathcal{S}^A_x
                \end{array}\right.,
\end{equation}
for some choice of output $a_x$ in every set $\mathcal{S}^A_x$. One can of course define an analogous operation over Bob's output. Any such operation will be called an output coarse graining and will be denoted $\mathcal{G}(\p)$. % \footnote{This operation can be regarded as the dual of the output liftings introduced in \cite{lift}.}.
Clearly, any such operation cannot create nonlocal resources out of local ones as $\mathcal{G}(\L)\in\l$ $\forall\L\in\l$ and every output coarse graining $\g$. This follows readily from the definition of $\l$ as given by Eqs.\ (\ref{local}) or (\ref{localdet}).
\begin{itemize}
\item[(P3)] $\p\succ\g(\p)$ $\forall\p\in\ns$ and every output coarse graining $\g$.
\end{itemize}
It should be stressed that (P3) cannot be included in (P2) because to map a non-vanishing probability to zero cannot be achieved by mixing with local behaviours (unless $\g(\p)\in\l$ which needs not necessarily happen). Thus, by mixing now $\g(\p)$ as in (P2), $\p$ can be compared to behaviours for which the mixing operation (P2) alone would have not sufficed. Moreover, as discussed above, the parties can then also transform to a mixture of different output coarse graining operations, i.\ e.\ $\p\succ\co(\p,\{\g_i(\p)\})$ for any output coarse graining operations $\{\g_i\}$.

The operations above allow the parties to merge different outputs. On the other hand, a party could also choose to split one outcome into two different ones. That is, Alice could enlarge artificially her set of outputs (with the possibility again to condition on the input) to $m_A+1$ by assigning a given output to the new output with a certain probability. That is, $P$ can be mapped to $P'$ such that for a given output $a$ and input $x$
\begin{eqnarray}
P'(ab|xy)=pP(ab|xy)\quad\forall b,y,\nonumber\\\label{unfold}
P'(m_A+1,b|xy)=(1-p)P(ab|xy)\quad\forall b,y,
\end{eqnarray}
with some $p\in[0,1]$ (while $P'=P$ otherwise). This kind of mappings and the analogous for Bob will be referred to as output unfolding operations and will be denoted by $\u(\p)$. As expected, it is easily checked that $\u(\L)\in\l$ $\forall\L\in\l$. As before, this allows to compare behaviours with different numbers of outputs. However, this can also be composed with some output coarse graining operation to transform back to a behaviour over the same setting, which is our primal interest. This translates to mappings of the form
\begin{eqnarray}
P'(a_1b|xy)=P(a_1b|xy)+pP(a_2b|xy)\quad\forall b,y,\nonumber\\\label{unfold2}
P'(a_2b|xy)=(1-p)P(a_2b|xy)\quad\forall b,y,
\end{eqnarray}
for some $p\in[0,1]$ and any given input $x$ and any two given outputs $a_1$ and $a_2$ (while $P'=P$ otherwise). This amounts to Alice deciding to regard output $a_2$ as $a_1$ with probability $p$ for input $x$. Fortunately, any such operation is already included above as this is nothing but a mixture of a behaviour and some output coarse graining of it. That is, the operation above reduces to $\p'=(1-p)\p+p\g(\p)$, where $\g$ is the output coarse graining operation merging $a_1$ and $a_2$ into $a_1$ for input $x$. This exhausts all possible operations over the outputs.

\subsubsection{Input operations}
It remains to analyze the transformations induced by acting locally on the input alphabets. First, one or both parties, can of course decide not to measure certain inputs, which simply eliminates some entries of $\p$. We call any such operation an input shortening and denote it by $\mathcal{S}(\p)$ \footnote{Output coarse graining and input shortening operations can also be regarded respectively as the dual of the output and input liftings introduced in \cite{lift}.}. Clearly, $\mathcal{S}(\L)\in\l$ $\forall\L\in\l$. Second, one or both parties could add an uncorrelated measurement input locally (which can be taken to be deterministic since they could later mix). That is, enlarge artificially the number of inputs of Alice to obtain a new behaviour $P'$ such that
\begin{equation}
P'(ab|xy)=\left\{\begin{array}{cc}
            P(b|y)\delta_{a,f(x)} & x>m_A  \\
            P(ab|xy) & x\leq m_A,
          \end{array}\right.
\end{equation}
with some function $f$ defined as in Eq.\ (\ref{deterministic}) over the enlarged input alphabet, and similarly for Bob. We call such mappings uncorrelated input largening and denote them by $\e_u(\p)$. To compare behaviours in the same setting one would need to compose these operations, i.\ e.\ replace one input by a deterministic uncorrelated input. Fortunately, this amounts to an output coarse graining operation $\g$ of all the outputs corresponding to that input and, therefore, they are already included in (P3). However, another possibility for a party to increase its input alphabet is to use some of the original inputs as new possible inputs, i.\ e.\ $P'(ab|m_A+1,y)=P(ab|x_1y)$ $\forall a,b,y$ for some input choice $x_1$. Transformations of this form will be referred to as correlated input largening and denoted by $\e_c$. Correlated and uncorrelated input largening operations will be both denoted by $\e$. It is again straightforward to verify that $\e(\L)\in\l$ $\forall\L\in\l$ as they map deterministic points (\ref{deterministic}) to deterministic points. Now, composing an input shortening operation with a correlated input largening amounts to transformations of the form
\begin{equation}\label{substitution}
P'(ab|x_2y)=P(ab|x_1y)\quad\forall a,b,y,
\end{equation}
for some given inputs $x_1$ and $x_2$ (while $P'=P$ otherwise). Such operations (and the equivalent for Bob) will be called input substitution and represented by $\k$.
\begin{itemize}
\item[(P4)] $\p\succ\k(\p)$ $\forall\p\in\ns$ and every input substitution $\k$.
\end{itemize}
As before, in more generality we have that $\p\succ\co(\p,\{\k_i(\p)\})$ for any input substitution operations $\{\k_i\}$. This reflects a party's possibility to use some input (with a certain probability) in exchange of another input.
%Last, a party could choose to perform a coarse graining of a subset of inputs. For instance, Alice could decide that there is no difference in choosing input 1 or 2 (corresponding to $p=1/2$ below). She could even choose to keep the labels or swap them with a certain probability, i.\ e.\ for some $p\in[0,1]$
%\begin{eqnarray}
%P'(ab|1y)=pP(ab|1y)+(1-p)P(ab|2y),\nonumber\\
%P'(ab|2y)=pP(ab|2y)+(1-p)P(ab|1y).
%\end{eqnarray}
%Fortunately again, this can be accounted by mixtures of $\p$ and some relabeling $\r(\p)$ and, hence, this is also already taken into account by the operations above. Thus, in summary, input transformations only add something when comparing behaviours over different settings.

\subsubsection{Wirings}

Finally, one would need to consider the operations known as wirings which consist in using the output obtained from a measurement on a behaviour as the input for a measurement on another one. However, this is only possible when several behaviours or several copies of a behaviour can be accessed jointly (or when several parties can act jointly). Since here we only deal with the single-copy manipulations of bipartite behaviours, we will not need to consider such operations \footnote{Wiring operations can be used to transform several copies of a given behaviour into one copy of a more nonlocal behaviour as in entanglement distillation (see \cite{distillation}).}.

\subsection{The ordering $\succ$ for no-signaling behaviours over the same setting}

As discussed in the previous subsection, properties (P1)-(P4) provide the basic structure of the set of allowed operations when restricted to transformations to behaviours over the same setting. If this is not the case the operations $\u$, $\mathcal{S}$ or $\e$ need to be considered in addition. To avoid such complication which does not add much conceptual difference, we will consider in the following only transformations over the same setting. As mentioned several times already composing and/or mixing of any of the operations given by (P1)-(P4) is also possible. Throughout the paper we will denote by $\o$ any operation which is a relabeling, an output coarse graining, an input substitution operation or a composition of them. Thus, we can then define the set of allowed operations, single-copy WCCPI \footnote{One could also use then CCPI to denote these operations.}, and its corresponding induced ordering in the following property:
\begin{itemize}
\item[(P)] Given $\p_1,\p_2\in\ns$, $\p_1\succ\p_2$ if $\p_2=p_0\L+\sum_{i=1}p_i\o_i(\p_1)$ for any convex weights $\{p_i\}$ ($i=0,1,\ldots$) and any $\L\in\l$.% and any operation $\o_i$ which is a relabeling, an output coarse graining or an input substitution operation or a composition of them.
\end{itemize}
Notice that the number of possible different operations $\{\o_i\}$ is finite and fixed by the setting. Notice also that the single-copy WCCPI operations as given by property (P) give rise to at least a partial ordering: (i) $\p\succ\p$ $\forall\p\in\ns$ (reflexivity), (ii) if for some $\p_1,\p_2,\p_3\in\ns$ it holds that $\p_1\succ\p_2$ and $\p_2\succ\p_3$, then $\p_1\succ\p_3$ (transitivity) and (iii) if for some $\p_1,\p_2\in\ns$ it holds that $\p_1\succ\p_2$ and $\p_2\succ\p_1$, then $\p_1$ and $\p_2$ are equally nonlocal (antisymmetry).

%\begin{itemize}
%\item Reflexivity: $\p\succ\p$ $\forall\p\in\ns$.
%
%\item Transitivity: If for some $\p_1,\p_2,\p_3\in\ns$ it holds that $\p_1\succ\p_2$ and $\p_2\succ\p_3$, then $\p_1\succ\p_3$.
%
%\item Antisymmetry: If for some $\p_1,\p_2\in\ns$ it holds that $\p_1\succ\p_2$ and $\p_2\succ\p_1$, then $\p_1$ and $\p_2$ are equally nonlocal.
%
%\end{itemize}

%Nonlocality should not increase under any input shortening or any input largening. However, it should also not increase when the parties ignore the possibility of measuring the new inputs created by the operation $\e$. Therefore, $\p$ and $\e(\p)$ should be equally nonlocal.
%\begin{itemize}
%\item[(P4)] $\p\succ\mathcal{S}(p)$ and $\p$ and $\e(\p)$ are equally nonlocal $\forall\p\in\ns$ and every input shortening $\mathcal{S}$ and input largening $\e$.
%\end{itemize}

Once the set of allowed operations has been mathematically specified in property (P), several questions arise. For instance, in the case of pure bipartite states it is known that there is a unique maximally entangled state as this is the only state that allows to obtain any other state by LOCC and cannot be obtained by LOCC from any other \cite{nielsen}. However, it has been recently shown that no such state exists in the multipartite case but a maximally entangled set of states needs to be considered instead \cite{us}. Similarly, we can ask what is the minimal set of behaviours that allows to obtain all other behaviours within the same setting through single-copy WCCPI operations. This question is answered in our case by noticing that $\ns$ is also a polytope (see e.\ g.\ \cite{nlreview}). Its vertices are the vertices of $\l$ together with some nonlocal vertices. Since mixing is an allowed operation, maximally nonlocal behaviours can then only correspond to the latter. These vertices have been characterized for the case of two inputs ($m_A=m_B=2$) and arbitrary outputs ($d=\min(d_A,d_B)$) in \cite{barrett}. They are all equivalent under relabelings as those given in (P1) to one of the behaviours $\{P_{PR}^{(k)}(ab|xy)\}_{k=2}^d$ whose only non-vanishing components are
\begin{equation}\label{prn2}
P_{PR}^{(k)}(aa|11)=P_{PR}^{(k)}(aa|21)=P_{PR}^{(k)}(aa|12)=P^{(k)}_{PR}(a,a+1|22)=1/k,
\end{equation}
where $a=1,2,\cdots,k$ and it should be understood $k+1=1$. This means that when $d=2$ there is a unique maximally nonlocal behaviour: $P_{PR}^{(2)}$. This is not the case when $d>2$ because one needs at least some\footnote{When $k$ divides $d$ it is possible to obtain $P_{PR}^{(k)}$ from $P_{PR}^{(d)}$ with some output coarse graining operation but otherwise this is not possible as follows from Eq.\ (\ref{prn2}). On the other hand, the fact that they are both vertices of the polytope $\ns$ forbids the possibility of transforming one into the other through mixing while in the two-input case input substitutions always map to local behaviours.} of the $P_{PR}^{(k)}$ with $k\leq d$ to be able to generate all no-signaling behaviours of the corresponding setting by single-copy WCCPI, thus leading to a maximally nonlocal set. A similar situation is found for the case of binary outputs ($d_A=d_B=2$) and arbitrary inputs, where the list of nonlocal vertices of $\ns$ can be found in \cite{JM}. Interestingly, many of these vertices can be obtained from others by input shortening, largening and/or substitution operations and it can be checked that when $m_A=m_B=3$ a unique one is actually enough to generate all the others within the same setting. However, leaving aside this case, the maximally nonlocal set for binary outputs has more than one element too\footnote{Interestingly, the existence of inequivalent vertices only holds in the single-copy case, as it is shown in \cite{JM} that, given enough copies of these extremal behaviours, interconversions are always possible.}.

Another natural question is whether $\succ$ induces a total order in $\ns$, i.\ e.\ whether given any $\p_1,\p_2\in\ns$ it always holds that either $\p_1\succ\p_2$ or $\p_2\succ\p_1$. However, this is not the case, i.\ e.\ $\succ$ is just a partial order. Intuitively, due to property (P2), totality could only arise if $\ns$ reduced to an interval. However, this is in fact a higher-dimensional polytope, and incomparable behaviours can be found (the inequivalent vertices stated above are for instance an example of this). The same thing happens in entanglement theory as there are entangled states which cannot be mapped into each other by LOCC in neither direction \cite{nielsen,us}.

Last, it would be interesting to characterize when a transformation from a no-signaling behaviour to another is possible, i.\ e.\ when $\p_1\succ\p_2$ holds. In the case of entanglement, this was done by Nielsen in \cite{nielsen} in the case of bipartite pure states (see \cite{us} for recent advances in the multipartite case). Although such a general characterization appears to be a difficult problem, the tools used to decide if a behaviour is local can also be used to approach this problem. This is because, fortunately, as given by property (P), to decide the behaviours that can be obtained from a given one $\p_1$ via single-copy WCCPI, it suffices to check over mixtures with a finite number of terms which are fixed by $\p_1$. It follows from there that $\p_1\succ\p_2$ if and only if (iff) $\p_2$ belongs to the polytope given by the convex hull of the vertices of $\l$, $\{\r_m(\p_1)\}$, $\{\g_n(\p_1)\}$, $\{\k_k(\p_1)\}$ and inequivalent compositions of these operations. Thus, $\p_1\succ\p_2$ iff
\begin{equation}
\p_2=\sum_jp_j\d_j+\sum_iq_i\o_i(\p_1),
\end{equation}
where all the weights above are nonnegative and $\sum_jp_j+\sum_iq_i=1$. This is mathematically analogous to the problem of deciding whether $\p\in\l$ and can also be solved by linear programming. If a solution is returned then we conclude that $\p_1\succ\p_2$, while if the problem turns out to be unfeasible this means that $\p_1\nsucc\p_2$. In this case a similar linear program can be written down to check whether $\p_2\succ\p_1$. If this problem is unfeasible too then it follows that $\p_1$ and $\p_2$ are incomparable. Notice, however, that these linear programming problems have a much larger number of variables than that of Eq.\ (\ref{localdet}) because relabelings, output coarse grainings and input substitutions of the given behaviour need to be taken into account. Thus, if the previous problem was already computationally unaffordable for large inputs, the situation can only be worse for the present problem.

%In the figure we depict an idealized representation of the sets $\l$ and $\ns$ together with an illustration of the ideas presented in this section.

\section{How much nonlocal is a behaviour?}

After being able to compare the nonlocality of different behaviours, the next natural step is to quantify how useful a given resource is. Once the set of allowed operations has been mathematically precisely defined as given in property (P), it is now straightforward to decide the fundamental requirements a nonlocality measure should meet. In the same way as the basic rule for entanglement measures is that they should not increase under LOCC operations, nonlocality measures should not increase under single-copy WCCPI manipulations. Therefore, a nonlocality measure should be a function mapping no-signaling behaviours to non-negative real numbers $N:\ns\rightarrow\mathbb{R}_0^+$ fulfilling the properties (N1) and (N2) below.
\begin{itemize}
\item[(N1)] N(\p)=0 if $\p\in\l$.
\end{itemize}
This fixes that local behaviours are not a resource and, hence, should contain zero nonlocality according to any measure. One could consider to write iff instead of if. However, the choice above reflects the possibility that some nonlocal behaviours could be useless for implementing a given task accounted by the operational meaning of a particular nonlocality measure $N$. Therefore, it seems possible that according to some nonlocality measure $N(\p)=0$ for some $\p\notin\l$. The analogous situation is considered in entanglement theory.

\begin{itemize}
\item[(N2)] If $\p_2=p_0\L+\sum_{i=1}p_i\o_i(\p_1)$ for some convex weights $\{p_i\}$ ($i=0,1,\ldots$) and some $\L\in\l$, then $N(\p_1)\geq N(\p_2)$.
\end{itemize}
As discussed above, this accounts for the fact that any nonlocality measure should respect the natural ordering $\succ$ induced by the allowed set of transformations. By manipulating a behaviour with single-copy WCCPI, nonlocality can never increase independently of the measure used. Notice that the fact that $N(\p)=N(\r(\p))$ is already included in (N2) as relabeling operations can be inverted by some other relabeling operation and we therefore have that $N(\p)\geq N(\r(\p))$ and $N(\r(\p))\geq N(\p)$.

Properties (N1) and (N2) are the fundamental requirements for any nonlocality measure which is defined over a given setting $\{m_A,m_B,d_A,d_B\}$ since all possible transformations among such behaviours are given by property (P). This is the situation we will be generally considering because one can work with measures that are only meaningful for a particular chosen setting\footnote{This happens as well in entanglement theory as there are entanglement measures which are only defined for a particular number of parties with given dimensionalities. For instance, the tangle \cite{tangle} is only defined for three-qubit systems.}. However, one could still consider more general measures which are defined for any possible setting. In this case, the measures must have the proper behaviour under the allowed operations that transform behaviours to behaviours in different settings. The following requirement must be, nevertheless, handled with care because the normalization of such a given measure could depend on the setting.
\begin{itemize}
\item[(N3)]  $N(\p)=N(\u(\p))$, $N(\p)\geq N(\mathcal{S}(\p))$ and $N(\p)=N(\e(\p))$ for every output unfolding $\u$, input shortening $\mathcal{S}$ and input largening $\e$.
\end{itemize}
This is because as discussed in the previous section the parties are allowed to implement these operations and, hence, nonlocality cannot increase under them. The equality cases follow from the fact that any $\u$ ($\e$) can be inverted by some $\g$ ($\mathcal{S}$), i.\ e.\ $\p=\mathcal{G}(\u(\p))$ ($\p=\mathcal{S}(\e(\p))$). Of course, as discussed in the previous section, the property must still hold when different operations of this form are mixed together and/or with operations as those given in (P), e.\ g.\ $N(\p)\geq N(\sum_ip_i\mathcal{S}_i(\p))$ for different input shortening operations $\{\mathcal{S}_i\}$ mapping $\p$ to the same setting with $\{p_i\}$ being convex weights, which is already accounted by the properties above because $\sum_ip_i\mathcal{S}_i(\p)=\mathcal{S}(\sum_ip_i\r_i(\p))$ for some fixed $\mathcal{S}$ and some relabeling operations $\{\r_i\}$.

The properties above provide the minimal requirements for a nonlocality measure. We are now in the position to develop such measures in a rigorous and systematic way. In the following we consider the most common nonlocality quantifiers introduced so far and analyze whether they qualify as proper measures from this point of view.

\subsection{Measures based on the violation of a Bell inequality}

A first common choice for the quantification of nonlocality is to use the amount of violation of a particular given Bell inequality. In other words, given some $\s\in\mathbb{R}^{d_Ad_Bm_Am_B}$ for which it is known that $\s\cdot\L\leq S$ $\forall\L\in\l$, the idea is to use $\s\cdot\p-S$ as a measure of the nonlocality of $\p$. Although the range of values this kind of measure might take depends strongly on the choice of Bell inequality (i.\ e.\ on the choice of $\s$) and on whether it corresponds to a facet or not, it appears as very intuitive that for every fixed $\s$, the larger $\s\cdot\p-S$ is, the more nonlocal $\p$ should be. Moreover, such measures have the advantage of being easily computed for every behaviour. It is clear that the value of $\s\cdot\p$ changes with the relabelings of $\p$, so when one speaks of the amount of violation of a Bell inequality, it is implicitly assumed that all possible relabelings are taken into account. That is, these measures are defined as
\begin{equation}\label{bellmeasure}
N_\s(\p)=\max\{\max_\r(\s\cdot\r(\p)-S),0\}.
\end{equation}
For instance, the standard choice when $m_A=m_B=d_A=d_B=2$ is the CHSH inequality which is usually expressed as
\begin{equation}\label{chsh}
\langle A_1B_1\rangle+\langle A_2B_1\rangle+\langle A_1B_2\rangle-\langle A_2B_2\rangle\leq2,
\end{equation}
where $\{A_i\}$ and $\{B_i\}$ represent the input measurements available for the parties and values $\pm1$ are assigned to the two possible outputs, i.\ e.\
\begin{equation}\label{expectchsh}
\langle A_iB_j\rangle=P(11|ij)+P(22|ij)-P(12|ij)-P(21|ij).
\end{equation}
However, it is always understood that $CHSH=\max_{i,j}|\langle A_1B_1\rangle+\langle A_2B_1\rangle+\langle A_1B_2\rangle+\langle A_2B_2\rangle-2\langle A_iB_j\rangle|$ to account for all possible relabelings. Thus, $N_{CHSH}(\p)=CHSH-2$.

Measures defined as in Eq.\ (\ref{bellmeasure}) are restricted to a given setting, so we do not need to discuss property (N3). They obviously satisfy (N1) and it is clear that they do not increase under mixing with local behaviours (P2) because of linearity. However, it turns out that they do not fulfill property (N2) because, for instance, they can increase under output coarse graining operations. To see this, consider the simplest possible setting $m_A=m_B=2$ and $d_A=d_B=3$ for which the CHSH is still a valid Bell inequality if we assign the value 0 to the third output (i.\ e.\ the expectation values are still given by Eq.\ (\ref{expectchsh})). It is simple to see that coarse graining the output 3 for both parties and all inputs into 1 or 2 should allow to increase the violation. Take, for example, $P_{PR}^{(3)}$ of Eq.\ (\ref{prn2}), for which $N_{CHSH}(P_{PR}^{(3)})=1/3$ but $N_{CHSH}(\g(P_{PR}^{(3)}))=4/3$ for the output coarse graining operation $\g$ described above. %Thus, Bell-inequality-violation based measures can only fulfill (N2) and be then proper nonlocality measures when Eq.\ (\ref{bellmeasure}) is modified to $\max_\g N_\s(\g(\p))$.
Similarly, examples can be found where these measures increase under input substitution operations (this can be seen in the simplest possible setting $m_A=m_B=3$ and $d_A=d_B=2$ using the $I_{3322}$ inequality and the intuition is to use $\k$ to replace uncorrelated inputs). Thus, this suggests to modify Eq.\ (\ref{bellmeasure}) to $N_\s(\p)=\max\{\max_\o(\s\cdot\o(\p)-S),0\}$.

\subsection{Nonlocal content}

The local content $P_L$ of a given behaviour $\p$ is given by
\begin{equation}
P_L(\p)=\max\{p\in[0,1]:\p=p\L+(1-p)\q,\;\L\in\l,\q\in\ns\}.
\end{equation}
The decomposition $p\L+(1-p)\q$ yielding the optimal values is known as the EPR2 decomposition of $\p$, from the name of the authors that introduced it \cite{epr2}. This allows to define the nonlocal content or EPR2 measure
\begin{equation}\label{epr2meas}
EPR2(\p)=1-P_L(\p).
\end{equation}
Interestingly, this measure is analogous to the best separable approximation measure in entanglement theory \cite{bsa}. Notice that this measure is defined independently of the setting and that $EPR2(\p)\in[0,1]$. The minimal value corresponds to local behaviours while the maximal is attained by the nonlocal vertices of the polytope $\ns$ such as $P_{PR}^{(d)}$ (and points that can only be written as a mixture of these vertices). The evaluation of this measure can also be addressed by linear programming \cite{fitzi} (although as in deciding if a behaviour is local or not, the procedure becomes inefficient as the number of inputs increases). $EPR2$ fulfills the conditions (N1)-(N3) presented above and it is therefore a valid nonlocality measure in this context. This is because, as it is easy to check, the local content cannot decrease by linear operations in $\ns$ that map local behaviours to local behaviours. Since single-copy WCCPI are clearly such kind of operations, the EPR2 measure cannot increase under the set of allowed transformations.

\subsection{Robustness measures}

In a similar spirit to the robustness of entanglement \cite{vidal}, one can define robustness measures of nonlocality considering how much local noise a nonlocal behaviour can tolerate before becoming local. The natural analogy is \cite{david,pankaj}
\begin{equation}\label{robustness}
R(\p)=\min\{p\in[0,1]:\exists\,\L\in\l,\;(1-p)\p+p\L\in\l\},
\end{equation}
which can be referred to as robustness of nonlocality. It holds then that $R(\L)=0$ $\forall\L\in\l$ while $R(\p)>0$ for any nonlocal behaviour $\p$. Again, it is clear that the robustness $R$ is a meaningful nonlocality measure in the sense considered here (i.\ e.\ satisfies (N1)-(N3)) because it cannot increase under linear operations in $\ns$ that map local behaviours to local behaviours, as follows easily from the definition \cite{pankaj}.

One can allow for more general robustness measures by considering the noise tolerance with respect to certain meaningful local behaviours. A particular physically relevant choice is given by the minimal threshold detection efficiency to close the detection loophole in a Bell experiment. This situation arises because in an experimental setting it can happen that the outcome of some measurements is sometimes not conclusive. A typical example is a no detection event in an experiment with photons. To deal with this situation one has to include an extra outcome for Alice and Bob, $a=b=0$, that accounts for this additional possibility. Thus, although one might had aimed to prepare the nonlocal behaviour $P$, observations actually correspond $\forall x,y$ to
\begin{equation}\label{noise}
P_\eta(ab|xy)=\left\{\begin{array}{cc}
                       \eta^2P(ab|xy) & a,b=1,\ldots,d_{A,B} \\
                       1-\eta^2 & a=b=0 \\
                       \eta(1-\eta)P(a|x) & b=0 \\
                       \eta(1-\eta)P(b|y) & a=0
                     \end{array}\right.,
\end{equation}
where $\eta\in[0,1]$ represents the detection efficiency of Alice and Bob's detectors, i.\ e.\ the probability with which measurements are carried out efficaciously and yield a conclusive output represented in the initial set $\{1,\ldots,d_{A,B}\}$. Notice that the transformation from $P$ to $P_\eta$ corresponds to an output unfolding operation $\u$ on $P$ and subsequent mappings of the form (\ref{unfold2}) in which a $(1-\eta)$ fraction of the probability of all outcomes for every possible input for Alice and Bob is transferred to the new outputs $a=b=0$. Thus, $\p\succ\p_\eta$ $\forall \eta$ and, therefore, while $\p\in\l$ implies $\p_\eta\in\l$ too, it can hold that $\p\notin\l$ but $\p_\eta\in\l$. This last phenomenon is usually referred to as the detection loophole. As a consequence, to be able to observe experimentally a nonlocal behaviour the detection efficiency $\eta$ must be sufficiently high so that $\p_\eta\notin\l$. One then defines the minimal threshold detection efficiency $\eta^*$ associated to a given behaviour $\p$ as
\begin{equation}\label{th}
\eta^*(\p)=\max\{\eta\in[0,1]:\,\p_\eta\in\l\}.
\end{equation}
This suggests the nonlocality measure
\begin{equation}\label{eff}
N_{eff}(\p)=1-\eta^*(\p).
\end{equation}
Notice that $N_{eff}(\L)=0$ $\forall \L\in\l$. The maximal value attainable by this measure depends on the setting \cite{mpth}. For instance, for the simplest setting $m_A=m_B=d_A=d_B=2$ it holds that $N_{eff}(\p)\in[0,1/3]$ \cite{mpth,branciard}. For obvious reasons, considerable attention has been paid to the the study of $\eta^*$ and intuition suggests its use as a nonlocality quantifier. It seems very natural to consider that the less detection efficiency a behaviour requires, the more nonlocal it should be. For instance, Eberhard's striking result \cite{eberhard} that weakly entangled states are more robust in this scenario than maximally entangled states is usually cited as an instance of the fact that more entanglement can correspond to less nonlocality. It seems therefore desirable to elucidate whether $N_{eff}$ can be regarded as a rigorous nonlocality measure. In the following we answer this question on the affirmative by proving that this measure satisfies the properties (N1)-(N3) we have postulated above for all nonlocality measures.

That (N1) is fulfilled is clear. To see that (N2) is satisfied too, consider the set
\begin{equation}\label{effset}
\l(\eta)=\{\p\in\ns:\,\p_{\eta'}\in\l\;\forall\eta'\leq\eta\}=\{\p\in\ns:\,N_{eff}(\p)\leq1-\eta\}.
\end{equation}
This set is clearly convex, as $\forall\p,\p'\in\l(\eta)$, $\lambda\p_{\eta'}+(1-\lambda)\p'_{\eta'}\in\l$ $\forall\eta'\leq\eta$ and $\forall\lambda\in[0,1]$ and, therefore, $\lambda\p+(1-\lambda)\p'\in\l(\eta)$ too (actually $\l(\eta)$ is also a polytope \cite{branciard}). Thus, to check whether (N2) is satisfied, it suffices to prove that $N_{eff}(\L)=0$ $\forall \L\in\l$, $N_{eff}(\g(\p))\leq N_{eff}(\p)$ for every output coarse graining operation $\g$, $N_{eff}(\k(\p))\leq N_{eff}(\p)$ for every input substitution operation $\k$ and $N_{eff}(\r(\p))=N_{eff}(\p)$ for every relabeling operation $\r$ (where these operations act on the original input and output alphabets). The first one has already been observed. The second follows from the fact that $(\g(\p))_\eta=\g(\p_\eta)$. Hence, if $\p_\eta\in\l$, then $(\g(\p))_\eta\in\l$ too and thus $\eta^*(\g(\p))\geq\eta^*(\p)$. The same reasoning shows that the property is fulfilled for $\k$. For the last one, we analogously observe that $(\r(\p))_\eta=\r(\p_\eta)$ together with the fact that $\p_\eta\in\l$ iff $\r(\p_\eta)\in\l$, which implies $\eta^*(\r(\p))=\eta^*(\p)$. Following exactly the same lines it can be readily checked that property (N3) also holds for $N_{eff}$ \footnote{Notice that it is in principle not obvious as in previous cases whether $N_{eff}$ does not increase under linear operations in $\ns$ that map local behaviours to local behaviours, $\Lambda$, as $(\Lambda(\p))_\eta=\Lambda(\p_\eta)$ does not need to hold in general for such operations.}.

Instead of considering extra outputs $a=b=0$, one can also associate the nonconclusive events to a particular valid outcome, say $a=b=1$. This, besides being experimentally convenient, is sometimes regarded as the optimal way to deal with nonconclusive measurements (see e.\ g.\ Sec.\ VII.B.1.b in \cite{nlreview}). Notice, however, that instead of using the behaviour $P_\eta$ given in Eq.\ (\ref{noise}), one models the observations in this case with the behaviour $\g(\p_\eta)$ where $\g$ is the output coarse graining operation that merges together the outputs 0 and 1 for all inputs. Thus, while $\p_\eta\in\l$ implies $\g(\p_\eta)\in\l$, it can still hold that $\p_\eta\notin\l$ but $\g(\p_\eta)\in\l$. It seems then surprising that one would prefer to model the observed behaviour using this possibility instead of the one given in Eq.\ (\ref{noise}). The solution to this apparent paradox is that to check whether the observed behaviour is nonlocal, the violation of a particular Bell inequality is tested. As discussed in Sec.\ 5.1, this is not a nonlocality measure and it can increase by output coarse graining operations. It can therefore happen that $\g(\p_\eta)$ passes the nonlocality test while $\p_\eta$ does not, even though the latter is more nonlocal than the former. Notice, however, that if more refined ways of detecting the nonlocality of the behaviour would be used, it must be always better to model it by $\p_\eta$. From this point of view, it comes as no surprise that the numerical results of \cite{wilms} show that $\eta^*$ is smaller when nonconclusive events are treated as a separated extra output.

%Since this local noise is fixed, this measure can only make sense for a given setting. Moreover, the noise tolerance has to be computed with respect to all possible relabelings of the local noise $\r(\L)$ as otherwise the measure would be clearly ill-defined. Thus, we define the robustness of nonlocality with respect to the local noise $\L$ as
%\begin{equation}
%R_\L(\p)=\min\{p\in[0,1]:\exists\,\r,\;(1-p)\p+p\r(\L)\in\l\}.
%\end{equation}
%As discussed above considering all possible relabelings $\r(\L)$ guarantees that $R_\L(\r(\p))=R_\L(\p)$.
%Although it is still evident that these measures vanish for every local behaviour while $1\geq R_\L(\p)>0$ for any nonlocal behaviour $\p$, it is no longer clear that (N2) is fulfilled. We prove that in the following. First we show that $R_\L(\g(\p))\leq R_\L(\p)$ for any output coarse graining operation. To see this, we use that the general robustness is a nonlocality measure and hence $R(\p)\geq R(\g(\p))$. Moreover, as it is a monotonic function of , there must exist some $\s$ such that $|\s\cdot\p|\geq|\s\cdot\g(\p)|$ with $|\s\cdot\p|>\max_{\L\in\l}|\s\cdot\L|=S$ and $|\s\cdot\L|\leq S$. Let $R_\L(\p)=r$, by definition this means that
%\begin{equation}
%|(1-r)\s\cdot\p+r\s\cdot\L|\leq S.
%\end{equation}
%Now, this together with the inequalities above implies that
%\begin{equation}
%|(1-r)\s\cdot\g(\p)+r\s\cdot\L|\leq S.
%\end{equation}

\subsection{Classical communication cost of simulation measures}

As mentioned many times already, in the DIQIP scenario the parties are not allowed to communicate during the measurement phase and they can only prepare local behaviours. An intuitive and operationally motivated measure of nonlocality is then to consider how much classical communication would need to be exchanged if this restriction was lifted in order to simulate a given nonlocal behaviour. To do this, one considers the signaling deterministic points of the form $\delta_{a,f(x,y)}\delta_{b,g(x,y)}$, where the outputs may now depend on the other's party choice of input through the functions $f$ and $g$ \cite{bacon}. These points can be grouped in the sets $\D_i$ according to the fact that $i$ bits of communication need to be exchanged by the parties in order to evaluate $f$ and/or $g$ ($\D_0$ corresponds to the local deterministic points (\ref{deterministic})). The deterministic points in  $\D_i$ are denoted $\{\d_{\lambda_i}\}$ and its cost by $C(\d_{\lambda_i})=i$ bits $\forall\lambda_i$. All the different decompositions of a given behaviour $\p=\sum_i\sum_{\lambda_i}p_{\lambda_i}\d_{\lambda_i}$ provide strategies on how to simulate it by a classical mixture of signaling deterministic events. Thus, the average communication cost \cite{pir} of a behaviour $\p$ is defined by the average cost of the optimal strategy, that is
\begin{equation}\label{avecomm}
\bar{C}(\p)=\min\{\sum_i\sum_{\lambda_i}p_{\lambda_i}C(\d_{\lambda_i}):\p=\sum_i\sum_{\lambda_i}p_{\lambda_i}\d_{\lambda_i}\}.
\end{equation}
On the other hand, the worst case communication \cite{bacon} is defined by the maximal amount of communication that needs to be used in the optimal protocol or in other words by
\begin{equation}\label{worcomm}
C_w(\p)=\min\{\max_iC(\d_{\lambda_i}):\p=\sum_i\sum_{\lambda_i}p_{\lambda_i}\d_{\lambda_i}\}.
\end{equation}
The definition of $\bar{C}$ is reminiscent of convex roof constructions of entanglement measures \cite{monotones} while that of $C_w$ of the Schmidt number \cite{schmidt} in entanglement theory. Notice that both $\bar{C}(\p)=0$ and $C_w(\p)=0$ iff $\p\in\l$. It is interesting to observe that these physically motivated measures are proper nonlocality quantifiers from the point of view taken in this paper, i.\ e.\ they satisfy properties (N1)-(N3). To see that (N2) holds one just needs to take into account the definition of the measures together with the fact that $\forall \d_{\lambda_i}$, $\r(\d_{\lambda_i})=\d_{\lambda'_i}$ for any relabeling operation $\r$ and that $\g(\d_{\lambda_i})=\d_{\lambda'_j}$ and $\k(\d_{\lambda_i})=\d_{\lambda''_k}$ with $j,k\leq i$ for any output coarse graining operation $\g$ and any input substitution $\k$. The analogous observation for $\u$, $\mathcal{S}$ and $\e$ shows that (N3) is fulfilled too.

\subsection{Statistical distance measures}

Another natural choice for nonlocality measures is to use some statistical distance to the set of local behaviours as this quantifies in some sense the support to the hypothesis that the observed data cannot be explained by a local behaviour. The most common measure to distinguish the behaviour $\p_1$ from $\p_2$ is given by the relative entropy or Kullback-Leibler distance
\begin{equation}\label{kl}
D(\p_1||\p_2)=\sum_{abxy}P_1(ab|xy)\log\left(\frac{P_1(ab|xy)}{P_2(ab|xy)}\right).
\end{equation}
This allows to define the measure \cite{dam}
\begin{equation}\label{relmeas}
D(\p)=\min\{D(\p||\L):\L\in\l\},
\end{equation}
which can be named relative entropy of nonlocality as this measure is completely analogous to the relative entropy of entanglement \cite{relent}. This measure has been studied as another instance of the fact that less entangled states can be more nonlocal \cite{toni}. It would be therefore interesting to check whether this measure qualifies as a proper nonlocality measure in our context. However, as we show below, $D$ does not satisfy property (N3). This might be not so surprising since, due to the definition (\ref{kl}), it seems likely that this measure increases when we enlarge the set of inputs. As discussed above, this could just be a matter of normalization which could be perhaps fixed by redefining (\ref{kl}) by dividing by the number of inputs $m_Am_B$. Thus, it is nevertheless interesting to check whether (N1)-(N2) are met, which are the relevant requirements when the setting is fixed. However, although it is obvious from the definition that property (N1) is satisfied (with an iff statement), I have not been able to prove that (N2) holds for this measure in full generality and it is thus left as an open problem. Technical details are given below.

To answer these questions, it is convenient to observe first that the relative entropy of nonlocality is convex, i.\ e.\ $D(\lambda\p_1+(1-\lambda)\p_2)\leq\lambda D(\p_1)+(1-\lambda)D(\p_2)$ $\forall\lambda\in[0,1]$ and $\forall\p_1,\p_2\in\ns$. This follows from the fact that the relative entropy is jointly convex on both input arguments \cite{cover}, i.\ e.\ if $\p_{1,2}^*$ denote the optimal local behaviours to compute $D(\p_{1,2})$ in Eq.\ (\ref{relmeas}), we have that
\begin{eqnarray}
D(\lambda\p_1+(1-\lambda)\p_2)\leq D(\lambda\p_1+(1-\lambda)\p_2||\lambda\p_1^*+(1-\lambda)\p_2^*)\nonumber\\
\leq\lambda D(\p_1||\p_1^*)+(1-\lambda) D(\p_2||\p_2^*)=\lambda D(\p_1)+(1-\lambda)D(\p_2).
\end{eqnarray}
Now, using the convexity of $D$, it suffices to check that $D(\r(\p))=D(\p)$ $\forall\r$, $D(\g(\p))\leq D(\p)$ $\forall\g$ and $D(\k(\p))\leq D(\p)$ $\forall\k$ in order to prove that property (N2) is met (the fact that $D(\L)=0$ $\forall\L\in\l$ has already been noticed). That the first one is true is a consequence of the fact that $D(\p_1||\p_2)=D(\r(\p_1)||\r(\p_2))$, i.\ e.\ for some $\L\in\l$% we have that
\begin{equation}
D(\p)=D(\p||\L)=D(\r(\p)||\r(\L))\geq D(\r(\p)),
\end{equation}
%and
%\begin{equation}
%D(\r(\p))=D(\r(\p)||\L')=D(\p||\r^{-1}(\L'))\geq D(\p),
%\end{equation}
where we have used that $\r(\L)\in\l$ $\forall\L\in\l$% and that the inverse of a relabeling operation is a relabeling operation too.
. Similarly, $D(\r(\p))\geq D(\p)$ has to hold too since the inverse of a relabeling operation is another relabeling operation. That $D(\g(\p))\leq D(\p)$ holds follows from the so-called log-sum inequality (see Theorem 2.7.1 in \cite{cover}), i.\ e.\
\begin{equation}
\sum_{i=1}^na_i\log\left(\frac{a_i}{b_i}\right)\geq\left(\sum_{i=1}^na_i\right)\log\left(\frac{\sum_{i=1}^na_i}{\sum_{i=1}^nb_i}\right)
\end{equation}
for any sets on nonnegative numbers $\{a_i\}_{i=1}^n$ and $\{b_i\}_{i=1}^n$. This is because this implies that
\begin{equation}
D(\p)=D(\p||\L)\geq D(\g(\p)||\g(\L))\geq D(\g(\p)),
\end{equation}
where we use that $\g(\L)\in\l$ if $\L\in\l$. However, we cannot repeat the same argumentation for $\k$ as $D(\p||\L)\geq D(\k(\p)||\k(\L))$ does not hold in general (e.\ g.\ take $\sum_{aby}P(ab|x_1y)\log(P(ab|x_1y)/L(ab|x_1y))>\sum_{aby}P(ab|x_2y)\log(P(ab|x_2y)/L(ab|x_2y))$ and let $\k$ be the substitution operation that replaces $x_2$ for $x_1$). This does, of course, not mean that $D(\k(\p))\leq D(\p)$ does not hold, which remains to be proved in order for (N2) to be true. Actually, it seems likely that a proper choice of $\L$ for which $\sum_{aby}P(ab|x_1y)\log(P(ab|x_1y)/L(ab|x_1y))=0$ could satisfy $D(\k(\p)||\L)\leq D(\p)$.% but I have not been able to find a proof.

Last, although $D(\p)=D(\u(\p))$ and $D(\p)\geq D(\mathcal{S}(\p))$ as follows along the same lines by noticing that $\forall\p_1,\p_2\in\ns$, $D(\p_1||\p_2)=D(\u(\p_1)||\u(\p_2))$ $\forall\u$ and $D(\p_1||\p_2)\geq D(\mathcal{S}(\p_1)||\mathcal{S}(\p_2))$ $\forall\mathcal{S}$ (the last one follows from the fact that $P(ab|xy)$ is a probability distribution for every fixed $x$ and $y$ and hence $\sum_{ab}P_1(ab|xy)\log(P_1(ab|xy)/P_2(ab|xy))\geq0$ $\forall x,y$), property (N3) is not satisfied because of input largening operations. To see this, consider an uncorrelated input largening $\e_u$ by party A. It is a simple exercise to notice that $D(\e_u(\p))\geq D(\p)$ with equality only if Bob's marginals for $\p$ and the optimal $\L$ yielding $D(\p)$ are equal for every input. However, there is no reason to believe that this requirement is met and actually Hardy's nonlocal behaviour provides a counterexample for this (see Tables XVI and XVII in \cite{dam}). It is also simple to construct examples for which $D(\e_c(\p))> D(\p)$.

%One could consider other measures of distinguishability to define other nonlocality quantifiers as in Eq.\ (\ref{relmeas}). Some possibilities are the trace or Kolmogorov distance $D(\p_1,\p_2)=\sum_{abxy}|P_1(ab|xy)-P_2(ab|xy)|/2$ or measures based on the fidelity $F(\p_1,\p_2)=\sum_{abxy}\sqrt{P_1(ab|xy)P_2(ab|xy)}$. As in the case of the relative entropy of nonlocality, it can be checked that such measures satisfy properties (N1)-(N3).

\section{Conclusions}

On the analogy of entanglement theory we have approached nonlocality from the point of view of a resource theory. In order to do this, one needs to consider the set of operations that are allowed in the DIQIP scenario, with which the parties cannot create nonlocality. Such transformations have been considered in previous literature and have been termed as WCCPI and play an analogous role to LOCC in entanglement theory. Here, we have analyzed and described in full mathematical detail what these operations amount to in single-copy transformations in the set of no-signaling behaviours. This allows to define a (partial) ordering in this set that establishes which behaviours are more nonlocal than others. Since the set of allowed transformations from a given behaviour reduces to a mixture of finitely many elements, it follows that the set of behaviours which are less nonlocal than a given one constitutes a convex polytope. This allows to use the same tools as for deciding if a behaviour is nonlocal to answer the question of whether a given behaviour is more nonlocal than another one. Furthermore, defining precisely the natural ordering in $\ns$ allows to establish the basic requirements for a nonlocality measure as any such measure must respect this ordering. This provides a systematic framework to develop nonlocality measures and we have thus studied the most relevant previously considered measures from this point of view. This has allowed to prove that operationally-motivated measures such as the threshold detector efficiency and the classical communication cost can be considered as rigorous nonlocality measures. On the other hand, intuitive and widely used quantifiers such as the amount of violation of a given Bell inequality can lead to inconsistencies. Although the relative entropy of nonlocality poses some problems when comparing behaviours over different settings, it could still be a meaningful measure when restricted to a fixed setting. However, a complete answer to this question has not been found and it is left for further research.

Some quantifiers were clearly proper nonlocality measures because it follows straightforwardly that they cannot increase under linear operations in $\ns$ mapping local behaviours to local behaviours. This is because, obviously, the set of allowed transformations must be a subset of this class operations. On the other hand, to determine whether some quantifiers are proper measures it was easier to work with a more precise description of the set of allowed operations. A natural question in this context is to determine whether the above mentioned inclusion is strict. In the case of entanglement theory, it is known that the mathematically simpler set of the so-called separable operations still maps separable states to separable states but it is, however, strictly more powerful than the physically relevant LOCC \cite{separable}. %Another interesting line of further research deals with the fact that to determine if a behaviour is nonlocal or not is computationally hard as the number of inputs increases and, thus, one should expect the same inefficiency problems for the evaluation of nonlocality measures which only vanish for local behaviours. %This is the case of the measures that have been considered here. Pulling further the analogy with entanglement theory, it seems it could be very useful to have a sufficient (but not necessary) condition for nonlocality giving a reasonable approximation of the set of local behaviours and which could be more efficiently computed as is the case for the positive partial transposition condition for entanglement \cite{peres}. This could furthermore allow to introduce an efficiently computable measure of nonlocality on the analogy of the negativity \cite{negativity}.
Finally, here we considered single-copy transformations. Since this only induces a partial order, many different nonlocality measures can be defined. It would be interesting to study if another scenario, e.\ g.\ (infinitely) many copies, singles out a particularly meaningful measure.

\textit{Note added.} Right after the completion of this manuscript I became aware of related work by Geller and Piani, which is also available as a preprint in \cite{marco}.

\begin{ack}

The author acknowledges financial support from the Spanish MINECO through
grants MTM 2010-21186-C02-02 and MTM2011-26912.
\end{ack}

\end{document}